\documentclass[%
 aip,
 amsmath,amssymb,
 reprint%
]{revtex4-1}

\usepackage{graphicx}
\usepackage{dcolumn}
\usepackage{bm}

\usepackage[utf8]{inputenc}
\usepackage[T1]{fontenc}
\usepackage{mathptmx}
\usepackage{etoolbox}
\usepackage{xcolor}
\usepackage[version=4]{mhchem}
\usepackage{upgreek}
\usepackage{hyperref}
\hypersetup{
    colorlinks=true,
    citecolor= teal,
    linkcolor= blue
    }

\makeatletter
\def\@email#1#2{%
 \endgroup
 \patchcmd{\titleblock@produce}
  {\frontmatter@RRAPformat}
  {\frontmatter@RRAPformat{\produce@RRAP{*#1\href{mailto:#2}{#2}}}\frontmatter@RRAPformat}
  {}{}
}%
\makeatother
\begin{document}

\preprint{AIP/123-QED}

\title{Accurate quantification of lattice temperature dynamics from ultrafast electron diffraction of single-crystal films using dynamical scattering simulations}

\author{Daniel B. Durham}
\email{dbdurham@berkeley.edu}
\affiliation{Department of Materials Science and Engineering, University of California, Berkeley, Berkeley, CA, United States}
\affiliation{National Center for Electron Microscopy, Molecular Foundry, Lawrence Berkeley National Laboratory, Berkeley, CA, United States}
 
\author{Colin Ophus}
\affiliation{National Center for Electron Microscopy, Molecular Foundry, Lawrence Berkeley National Laboratory, Berkeley, CA, United States}

\author{Khalid M. Siddiqui}
\affiliation{Materials Science Division, Lawrence Berkeley National Laboratory, Berkeley, CA, United States}

\author{Andrew M. Minor}
\affiliation{Department of Materials Science and Engineering, University of California, Berkeley, Berkeley, CA, United States}
\affiliation{National Center for Electron Microscopy, Molecular Foundry, Lawrence Berkeley National Laboratory, Berkeley, CA, United States}

\author{Daniele Filippetto}
\affiliation{Accelerator Technology and Applied Physics Division, Lawrence Berkeley National Laboratory, Berkeley, CA, United States}

\date{\today}

\begin{abstract}

In ultrafast electron diffraction (UED) experiments, accurate retrieval of time-resolved structural parameters such as atomic coordinates and thermal displacement parameters requires an accurate scattering model. Unfortunately, kinematical models are often inaccurate even for relativistic electron probes, especially for dense, oriented single crystals where strong channeling and multiple scattering effects are present. This article introduces and demonstrates dynamical scattering models tailored for quantitative analysis of UED experiments performed on single-crystal films. As a case study, we examine ultrafast laser heating of single-crystal gold films. Comparison of kinematical and dynamical models reveals the strong effects of dynamical scattering within nm-scale films and their dependence on sample topography and probe kinetic energy. Applied to UED experiments on an 11 nm thick film using 750 keV electron probe pulses, the dynamical models provide a tenfold improvement over a comparable kinematical model in matching the measured UED patterns. Also, the retrieved lattice temperature rise is in very good agreement with predictions based on previously measured optical constants of gold, whereas fitting the Debye-Waller factor retrieves values that are more than three times lower. Altogether, these results show the importance of dynamical scattering theory for quantitative analysis of UED, and demonstrate models that can be practically applied to single-crystal materials and heterostructures.

\end{abstract}

\maketitle

\section{\label{sec:Intro}Introduction}

Ultrafast electron diffraction (UED) has emerged as a powerful tool for structural dynamics research, allowing to record excited-state atomic structure evolution with sub-picosecond temporal resolution.\cite{sciaini2011UEDreview,luiten2010RFcompressedFED,weathersby2015SLACUED,zhu2015BNLUED,qi2020UEDat50fs} Recently, UED has been applied to study diverse phenomena including nonequilibrium phases and transformations in quantum materials,\cite{zong2021quantummaterialUEDReview} formation of warm dense matter,\cite{slac2021warmdensematterUED} and electron-phonon coupling mechanisms,\cite{durr2021UEDSReview} to name a few. The broad utility of the technique lies in the sensitivity of diffraction signals to many structural features such as crystal structure, lattice strain, atomic coordinates, and thermal displacement parameters.\cite{humphreys1979scattering}

While analysis of UED data has often examined the evolution of diffraction signals themselves, additional important information can be gained by quantitatively retrieving the underlying structural parameters. For instance, time-resolved lattice temperature informs the role of heat flow in the observed dynamics, and can be used to determine other quantities such as electron-lattice coupling constants,\cite{qiu1992short,jiang2005improved} time constants for defect and interfacial scattering,\cite{sokolowski2015BiUED} and thermal conductivity within or between layers.\cite{krenzer2006thermalboundaryconductance} Also, accurate time-dependent structure retrieval may reveal new transient phases or finer structural details of metastable phases.

However, many solid-state UED studies are performed on dense, oriented single-crystal films,\cite{TaTe2UED,zong2021quantummaterialUEDReview,durr2021UEDSReview,slac2021warmdensematterUED,sciaini2011UEDreview} which pose challenges for quantitative structural retrieval. Single crystals are often chosen because they provide numerous advantages for UED, including strong peaks associated with individual diffraction orders, access to diffuse scattering between the peaks to study phonon population dynamics,\cite{durr2021UEDSReview,siwick2018UEDSGraphite,cotret2019UEDS} lack of grain boundary scattering which factors into polycrystalline film dynamics,\cite{hostetler1999epcouplinggraindep} and opportunity for polarization-dependent study and control.\cite{ernstorfer2020anisotropic} In addition, some emerging materials are primarily available as single crystals.\cite{pistawala2022crystalgrowthreview,samarth2017quantummaterialssynthesis} On the other hand, the high scattering cross section for electron probes often leads to multiple scattering, and in single crystals, this is compounded by electron channeling down atomic columns.\cite{kambe1974channelingsims} These effects are especially strong when probing dense, inorganic solids along high-symmetry zone axes. Despite the reduced cross section for relativistic electron probes, such effects have still been observed in some experiments even at MeV-scale energies.\cite{TaTe2UED} When these effects dominate, kinematical approximations are no longer valid and modeling the complete ``dynamical'' diffraction process is required to match the diffraction signals. 

\begin{figure*}
    \centering
    \includegraphics[width=17.5cm]{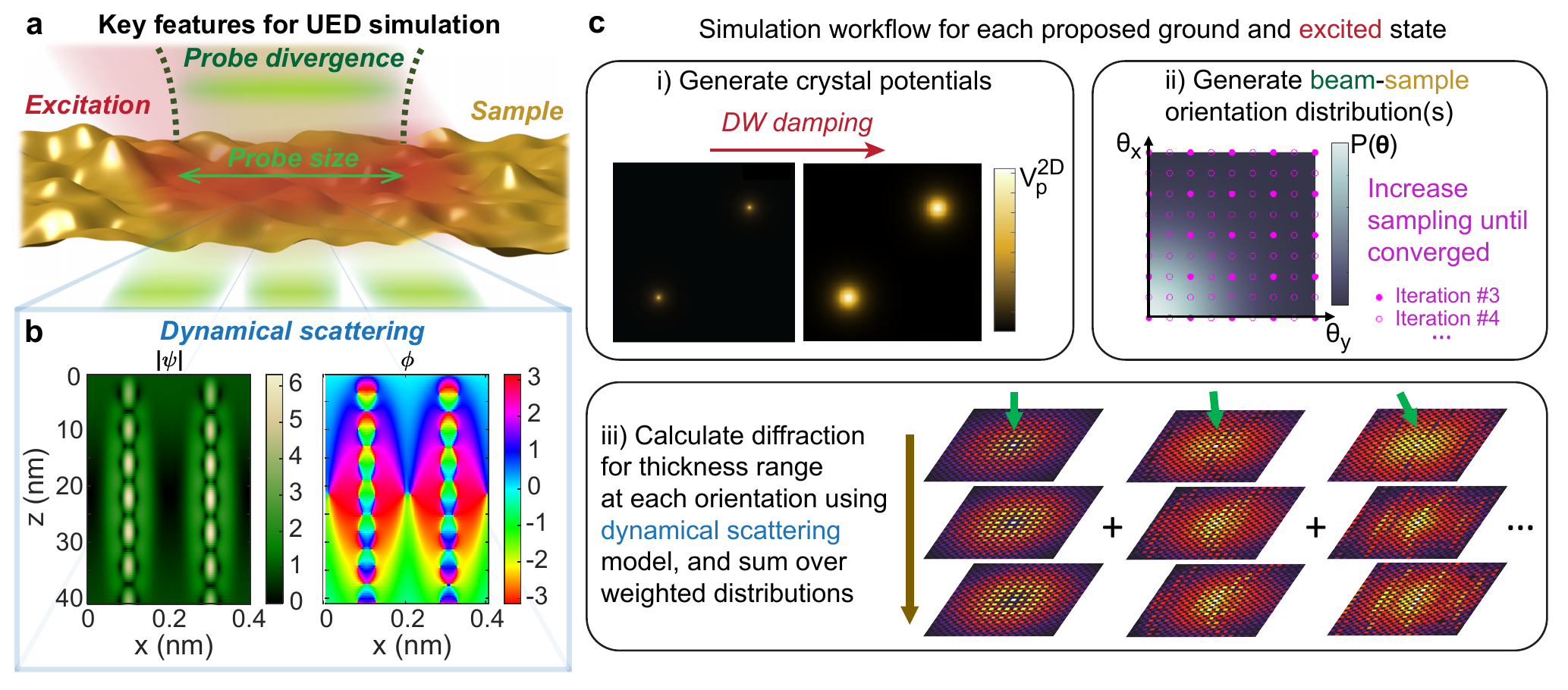}
    \caption{Modeling ultrafast electron diffraction signals from single crystal foils using dynamical scattering theory. a) Illustration of a UED experiment on a freestanding crystal foil, highlighting some key features to include for accurate models. b) Channeling plots showing an example of strong dynamical scattering effects. These were simulated using the mutlislice method for a 750 keV electron wave through gold [001] with mean square displacements of $\overline{u^2} = 0.024$ \AA$^2$. The amplitude, |$\psi$| (relative to incident amplitude of 1), and phase, $\phi$ (in radians), across a single gold unit cell through a 40 nm thick crystal are shown. c) Graphical summary of the procedure for UED simulations performed in this work. Images in step (i) are of a projected potential ($V^{\rm{2D}}_{\rm{p}}$) slice for gold [001] before and after applying Debye-Waller (DW) damping to account for thermal displacements: $\overline{u^2} = 0.024$ \AA$^2$ ($T$ = 300 K) is shown as an example. Note these images are each normalized by the maximum value for clearer visualization. Image in step (ii) shows a generated 2D Gaussian orientation distribution, which is sampled using increasingly dense grids like shown by the pink dots until convergence. Diagram in step (iii) illustrates computation of a thickness-dependent stack of diffraction patterns at each orientation. Diffraction patterns for beam-sample angles of 0 mrad (left), 50 mrad (center), and 100 mrad (right) at 1 (top), 6 (middle), and 11 (bottom) unit cells are shown as examples.}
    \label{fig:overview}
\end{figure*}

Suitable dynamical scattering models for UED of single-crystal films are needed for accurate quantification, but efforts to develop such models have so far been limited. Multiple scattering theory has been applied to UED of crystal surfaces in the reflection geometry\cite{schafer2011structural} and to strain wave imaging in UEM.\cite{du2020imaging} As for UED in the transmission geometry, a few studies have invoked Bloch wave eigenvalue (also called ``N-beam'') \cite{vallejo2018SiUED,jianqili2017dynamicalTaSeTe,li2019TaSeTe,bnl2022LMOUED} or multislice\cite{maxson2022kevmicroUED} simulations to improve fitting to measured signals as well as accuracy of retrieved lattice temperature and phonon dynamics. However, many of these examples added empirical functions to quantitatively match the diffraction signals, indicating room for further improvement in the underlying models. For instance, some factors that are not always considered for TEM simulations become important for UED, including lattice temperature, partial coherence of the probe, and sample topography (averaged over the typical $\rm{\mu}$m to mm UED probe size).

In this article, we demonstrate dynamical scattering models that are suitable for matching UED signals from single-crystal films and retrieving the lattice temperature dynamics. We first describe the computational approaches used, including both a multislice and a Bloch wave method, and introduce adaptations to account for key physical parameters. We then illustrate the role of dynamical scattering in UED of single-crystal films by comparing static and temperature-dependent diffraction signals calculated using kinematical and dynamical models for gold films of varying thickness and rippling as well as varying electron probe energy. Finally, we apply these models to analyze relativistic UED measurements of single-crystal gold films recorded at the High Repetition-rate Electron Scattering (HiRES) beamline at Lawrence Berkeley National Laboratory.\cite{filippetto2016HiRES} We show quantitative matching of static UED patterns, obtaining a factor of 10 improvement from the dynamical models over the kinematical model and achieving an R factor of 2\%. We then demonstrate lattice temperature retrieval, showing that dynamical scattering models provide good agreement with expectations based on the known optical properties of gold, whereas the kinematical model underestimates the expected temperature rise by nearly three times.

\section{UED simulation methods}

The simulation approach in this work was developed to account for several important factors in UED experiments of single-crystal foils, which are illustrated in Figure~\ref{fig:overview}a. An excitation such as a laser pulse triggers a dynamic process, leading to numerous potential excited states. The diffraction signals for each state are an average over the $\rm{\mu}$m to mm-scale probed region, which can consist of a wide beam-sample orientation distribution due to the sample topography and the probe beam divergence. Also, the electron probe can experience multiple scattering and channeling effects as described by dynamical scattering theory: for example, Figure~\ref{fig:overview}b shows a simulation of the amplitude and phase of the envelope of a 750 keV electron wave propagating through a 40 nm thick gold crystal oriented along [001]. Within just a few nm, the amplitude and phase become highly non-uniform and show complex variation with thickness. 

Our procedure is summarized in Figure~\ref{fig:overview}c. For each proposed ground and excited state, we generate the electrostatic potentials for the crystal, generate possible beam-sample orientation distributions, simulate the thickness-dependent diffraction patterns for the sampled orientations, and then compute weighted sums of the patterns according to the proposed distribution. In this section, we will first discuss the underlying scattering models used, and then describe how the beam-sample orientation distributions and excited states (in this case, crystals with varying lattice temperature) were incorporated. We note here that the Bloch wave and multislice methods give nearly equivalent results over the parameter ranges studied, so we use these methods interchangeably throughout the article (see the \hyperref[appendix]{Appendix}).

\subsection{Models for diffraction from a single-crystal film}

\subsubsection{Kinematical scattering}
\label{sec:kinscatter}

Formulae for calculating diffraction signals in the kinematical approximation have been described in detail elsewhere,\cite{humphreys1979scattering,Kirkland_2010} so we only elaborate on the details specific to our approach here.
In this work, the ``weak phase object'' (also called ``Moliere'') approximation is used to compute the atomic scattering factors, $f_{\mathrm{e},j}$, from parameterized atomic electrostatic potentials, $V_{\mathrm{a}}$, as calculated by Kirkland using a relativistic Hartree-Fock program:\cite{Kirkland_2010,grant1970relativistic}
\begin{equation}
    f_\mathrm{e}(q) = 
    \frac{2\pi i}{\lambda} \int_{0}^{\infty} J_{0}(2\pi q r) \left[1-\exp\left(i\sigma_{\mathrm{e}} \int V_{\mathrm{a}}(r,z) dz\right)\right] r dr 
\end{equation}
where $J_0$ is the zeroth order Bessel function, $q$ is the reciprocal space distance, and $r$ is the real space distance. We use the relativistic electron interaction parameter, $\sigma_e$, defined as:
\begin{equation}
    \sigma_{\mathrm{e}} =
    \frac{2\pi}{\lambda E_0} 
    \left(
    \frac{m_0 c^2+eE_0}{2m_0 c^2+eE_0}
    \right)
    \label{eq:sigma}
\end{equation}
For de Broglie wavelength $\lambda$, kinetic energy $E_0$, electron rest mass $m_0$, speed of light $c$, and electron charge $e$. 

We note that we did not use absorptive electrostatic potentials in this work because, in our case, they mostly remove electrons that should in fact remain included. The dominant contribution to absorptive potentials is typically thermal diffuse scattering (TDS).\cite{weickenmeier1991computation} However, the angle spread of the UED probe in our case is large enough that much of the TDS remains in the measured diffraction peaks and should not be removed from the simulations. Future work could examine modifications to absorptive potentials for such cases, though in our case we expect it would only provide minor corrections.

The scattering factors are then used to compute the structure factor for the periodic crystal, $F_{hkl}$, for each diffracted beam at reciprocal lattice vector $\mathbf{g}_{hkl}$ with Miller indices $h$, $k$, and $l$:
\begin{equation}
    F_{hkl} = \sum f_{\rm{e},\it{j}}(|\mathbf{g}_{hkl}|) \exp\left(2\pi i \mathbf{g}_{hkl}\cdot\mathbf{r}_j\right)
\end{equation}
Applying the shape factor for a thin film gives the diffracted intensity as a function of film thickness, $t$:\cite{cowley1992EDbook}
\begin{equation}
    I_{hkl} = \frac{\sin^2{(\pi s_{hkl} t})}{(s_{hkl} \xi_{hkl})^2}
\end{equation}
where $s_{hkl}$ is the excitation error
and $\xi_{hkl}$ is the extinction distance:
\begin{equation}
    \xi_{hkl} = \frac{\pi V_{\mathrm{cell}} \cos(\beta)}{\lambda |F_{hkl}|}
    \label{eq:extDist}
\end{equation}
where $V_{\mathrm{cell}}$ is the unit cell volume and $\beta$ is the angle between the beam and the surface normal.
As shown in Figure~\ref{fig:overview}b, the ``weak phase object'' approximation which underlies this method can be rapidly violated in dense, oriented crystals even at relativistic beam energies. Along atomic columns, the phase is strongly disturbed, leading also to strong modification of the amplitude envelope. This will be examined further in Section~\ref{sec:Sims}.

\subsubsection{Bloch waves}

For thicker specimens where kinematical approximations no longer hold, scattering patterns can instead be calculated by solving the Schr\"{o}dinger equation for the electron wave passing through the specimen. The Bloch wave eigenvalue solution is convenient for crystals. The electron wave and specimen potential are decomposed into Fourier components and a matrix equation is derived by which the electron wave components (and hence diffracted intensities) can be computed at varying distances through the crystal. 

The derivation of the matrix equation and the approximations used here are given in Ref.~\citenum{Kirkland_2010}. The computational procedure is to first calculate the Fourier components of the scattering potential, $U_{hkl}$:
\begin{equation}
    U_{hkl} = \frac{\sigma_\mathrm{e}}{\pi\lambda}V_{hkl} = \frac{\sigma_\mathrm{e}}{\pi\lambda}\frac{h^2}{2 \pi m_0 e V_{\mathrm{cell}}}F_{hkl}^{*}
\end{equation}
where h is Planck's constant, $F_{hkl}^{*}$ is the complex conjugate of $F_{hkl}$ computed using the first Born approximation,\cite{Kirkland_2010} and the other symbols are defined above. In addition, the excitation errors $s_{hkl}$ are calculated.

Then, a subset of the Fourier components is selected to be included in the simulation: namely, those with non-zero $U_{hkl}$, in-plane reciprocal space distance $q_{\mathrm{xy}} < q_{\mathrm{xy,max}}$, and $s_{hkl} < s_{\mathrm{max}}$. The thresholds $q_{\mathrm{xy,max}}$ and $s_{\mathrm{max}}$ are set such that the diffracted beam intensities of interest are converged (see subsection~\ref{subsec:convergence}). For this work, $q_{\mathrm{xy,max}}=4.5$~\AA$^{-1}$~and $s_{\mathrm{max}}=0.1$~\AA$^{-1}$. Note that Fourier components beyond those of the signals of interest are included since the diffracted beams interact with each other.

The scattering potential components and excitation errors are used to build the matrix $\mathbf{A}$ in which:
\begin{equation*}
    [a_{ij}] =  \begin{cases} 
    \text{$2k_0 s_{\mathbf{g}_j}$} & \text{$i=j$}\\
    \text{$U_{\mathbf{g}_j-\mathbf{g}_i}$} & \text{$i \neq j$}
    \end{cases}
 \end{equation*}
where $k_0$ is the incident electron wave vector and $i,j$ are indices for the Fourier components included in the simulation. Notably, computing $U_{\mathbf{g}_j-\mathbf{g}_i}$ typically requires computing $U$ at scattering vectors outside of the Fourier components selected for the simulation. The eigenvalues, $2k_{0,\mathrm{z}}\gamma_j$, and eigenvectors, $C_j$, of this $\mathbf{A}$ matrix can then be used to compute the electron wave $\psi$ in terms of the chosen Fourier components as a function of depth, z:
\begin{equation}
    \psi(z) = \mathbf{C}
    \left[
    \exp\left(2\pi i \gamma_j z\right)
    \right]
    \mathbf{C^{-1}}\psi(z=0)
\end{equation}
where $\mathbf{C}$ is a matrix with eigenvectors $C_j$ as the columns.
The diffracted beam intensities are then:
\begin{equation}
    I_{\mathbf{g}}(z) = |\psi_{\mathbf{g}}(z)|^{2}
\end{equation}

\subsubsection{Multislice}

Another solution to the aforementioned Schr\"{o}dinger equation is to divide the specimen into a series of 2D projected potential slices, interacting the wave with each slice and then propagating to the next slice. This multislice approach is widely used for electron microscopy image simulation and has been described in detail elsewhere,\cite{Kirkland_2010,Ophus_2017} so it will only be described briefly here. 

In this work, the envelope of the electron wave function, $\psi$, is initialized as a plane wave. It is then advanced through each slice $j$ with thickness $\Delta$t by applying two operators sequentially. First, the interaction operator is applied in real space:
\begin{equation}
    \psi_{j+1}(\mathbf{r})=
    \psi_{j}(\mathbf{r})
    \exp\left(i\sigma_{\mathrm{e}} V^{\mathrm{2D}}_{\rm{p}}(\mathbf{r})\right)
\end{equation}
Here, $V^{\rm{2D}}_{\rm{p}}(\mathbf{r})$ is the projected electrostatic potential within the slice, computed using the parameterized atomic potentials as determined by Kirkland, \cite{Kirkland_2010} and $\sigma_{\mathrm{e}}$ is the relativistic interaction parameter as defined in Equation~\ref{eq:sigma}. Second, a propagation operator is applied in reciprocal space:
\begin{equation}
    \psi_{j+1}(\mathbf{q})=
    \psi_{j}(\mathbf{q})\exp\left(-i\pi\lambda|\mathbf{q}|^{2}\Delta t\right)
\end{equation}

Finally, the diffraction signals are obtained from the Fourier transform of the exiting envelope:
\begin{equation}
    I(\mathbf{q}) = \left|\mathcal{F}(\psi(\mathbf{r}))\right|^{2} = |\psi(\mathbf{q})|^{2}
\end{equation}

Since the simulations in this work are oriented along the [001] zone axis of a cubic crystal, we choose the slices to be equally spaced, each containing one layer of atoms. Also, the simulation cell is a single unit cell in the models used here. We note that modeling thermal diffuse scattering, such as by using the frozen phonon approach, requires larger simulation cells.\cite{lamoen2008TDSDWMultislice} The image size was 256 px x 256 px, chosen to achieve convergence (see Section~\ref{subsec:convergence})

\subsection{Incorporating thermal motions}
\label{subsec:thermalMotions}

For this work, lattice temperature is incorporated by applying Debye-Waller damping to the projected potentials. This approximation models the electron beam traveling through a time-averaged electrostatic potential, and has been shown to account for the influence of thermal motions on the coherent Bragg diffraction peaks.\cite{lamoen2008TDSDWMultislice} As explained in section~\ref{sec:kinscatter}, we expect thermal diffuse scattering (TDS) to have only minor effects on the measured peak intensities in our case and so it is not included here. Where TDS plays an important role, diffraction simulations with finer q resolution over multiple "frozen phonon" configurations could be averaged together,\cite{lamoen2008TDSDWMultislice} though at much greater computational expense. 

Atoms moving randomly and independently with an RMS displacement of $u_{\mathrm{RMS}}$ along each dimension forms a 2D gaussian distribution of positions in the plane:
\begin{equation}
    f_{\mathrm{th},u_{\mathrm{RMS}}}(r) = 
    \frac{1}{2\pi u_{\mathrm{RMS}}^2} \exp
    \left(
    -\frac{r^2}{2 u_{\mathrm{RMS}}^2}
    \right)
\end{equation}
This distribution is convolved with the electrostatic potential in real space, effectively acting as a Gaussian filter. This filter in Fourier space is:
\begin{equation}
    f_{\mathrm{th},u_{\mathrm{RMS}}}(q) = \exp\left(-2\pi^2 u_{\mathrm{RMS}}^2 q^{2}\right)
\end{equation}
This approach is readily generalized to anisotropic or anharmonic thermal motions by applying the corresponding two-dimensional filter.

In kinematical theory, this filter is applied to the diffracted intensities as a Debye-Waller Factor (DWF):
\begin{multline}
    \frac{I_{hkl}(u_{\mathrm{RMS}})}{I_{hkl}(u_{\mathrm{RMS}}=0)} =
    \mathrm{DWF}(q_{hkl}) =
    f_{\mathrm{th},u_{\mathrm{RMS}}}(q_{hkl})^2 \\ 
    = \exp\left(-4\pi^2 u_{\mathrm{RMS}}^2 q_{hkl}^{2}\right)
\end{multline}

When kinematical approximations are valid, this allows extraction of a change in RMS displacements, $\Delta u_{\mathrm{RMS}}$, from measurements of diffraction intensities for a set of diffraction orders through linear regression using the form:
\begin{equation}
    -\rm{log}
    \left(
    \frac{I_{\mathit{hkl},2}}{I_{\mathit{hkl},1}}
    \right) 
    = 4\pi^2 \Delta u_{RMS}^2 q_{\mathit{hkl}}^{2}
\label{eq:DWFitting}
\end{equation}

Meanwhile, in the Bloch wave approach, this filter is applied to the scattering potential components:
\begin{equation}
    \frac{U_{hkl}(u_{\mathrm{RMS}})}{U_{hkl}(u_{\mathrm{RMS}}=0)} = f_{th,u_{\rm{RMS}}}(|\mathbf{g}_{hkl}|)
\end{equation}

Whereas in the multislice approach, this filter is applied to the projected potential slices:

\begin{equation}
    \frac{V^{\rm{2D}}_{\rm{p}}(\mathbf{q},u_{\rm{RMS}})}{V^{\rm{2D}}_{\rm{p}}(\mathbf{q},u_{\rm{RMS}}=0)} = f_{\rm{th},\it{u}_{\rm{RMS}}}(|\mathbf{q}|)
\end{equation}
An example of this is shown for the first slice from a gold [001] unit cell at a temperature of 300 K in Figure~\ref{fig:overview}c.i.

In the dynamical scattering models, the relationship between $\Delta u_{\rm{RMS}}$ and changes in diffracted intensities are more complex and depend greatly on both intrinsic and extrinsic sample properties such as material, orientation, thickness, sample topography, as will be illustrated in Section~\ref{sec:Sims}. A simple analytical formula like Equation~\ref{eq:DWFitting} does not generally exist for dynamical scattering; instead, $\Delta u_{\rm{RMS}}$ can be determined by minimizing the least squares error between simulated and measured diffraction intensity changes. 

\subsection{Orientation averaging}
\label{subsec:avgrippling}

In many UED experiments, a distribution of beam-sample orientations is sampled simultaneously due to two factors: Sample rippling within the large (often mm-scale) lateral probe size, and angular spread of the beam due to partial coherence. We model this by computing and incoherently summing the diffraction signals over a distribution of tilt angles, ie. by computing the following weighted integral over the 2D orientation space, $A$:
\begin{equation}
    I_{hkl,\rm{avg}} = \iint_A p(\theta_{\rm{x}},\theta_{\rm{y}})I_{hkl}(\theta_{\rm{x}},\theta_{\rm{y}}) \,d\theta_{\rm{x}}\,d\theta_{\rm{y}}
\end{equation}
Where $\theta_{\rm{x}}$ and $\theta_{\rm{y}}$ are the horizontal and vertical tilt angles, and $p(\theta_{\rm{x}},\theta_{\rm{y}})$ is the distribution of orientations. A similar approach has been used to model precession electron diffraction (PED),\cite{midgleyPEDMultislice} and this approach has been employed by others to account for UED beam divergence.\cite{slac2021SiUEDRockingCurves} Notably, this is a 2D integral and has significant contributions from near-zone orientations, so it will be more expensive to compute and more sensitive to the sampling than 1D integrals for PED which may avoid sampling near the zone axis.

In these calculations, $p(\theta_{\rm{x}},\theta_{\rm{y}})$ is approximated as a circularly symmetric Gaussian distribution, like illustrated in Figure~\ref{fig:overview}c.ii. This distribution can be considered as the convolution of the beam angular spread with the sample rippling such that the RMS tilt spread, $\sigma_{\uptheta}$, is given by:
\begin{equation}
    \sigma_{\uptheta} = \sqrt{\sigma_{\uptheta,\rm{sample}}^2 + \sigma_{\uptheta,\rm{probe}}^2 }
\end{equation}
$\sigma_{\uptheta,\rm{probe}}$ is typically on the order of 1 mrad or less for transmission UED setups,\cite{slac2021SiUEDRockingCurves} and $\sigma_{\uptheta,\rm{sample}}$ depends on the sample preparation but can be as much as tens of mrad in some cases.\cite{TaTe2UED} 

Since the calculations in this work are for a zone axis with four-fold symmetry, gold [001], the sampling grid for a given $\sigma_{\uptheta}$ just spans the positive quadrant in orientation space from 0 to 3$\sigma_{\uptheta}$, and then four-fold rotational averaging is applied to account for the other quadrants. Points located more than 3$\sigma_{\uptheta}$ from the center are set to zero to maintain circular symmetry. In this way, $\approx$ 99\% of the Gaussian volume is sampled. 

We compute the integral over the orientation space using an iterative 2D trapezoidal quadrature algorithm.\cite{press2007numericalrecipes} On the first iteration, a square sampling grid is initialized with just four samples: one at each corner. Then, on each successive iteration, points are added to complete a sampling grid with half the spacing in each dimension. A running integral is computed by adding 3/4 of the newly integrated points to 1/4 of the previous integral value. The sampling points for iterations 3 and 4 of this procedure are shown in Figure~\ref{fig:overview}c.ii as an example. 

The computations in this work sampled a 480 mrad $\times$ 480 mrad tilt range to examine $\sigma_{\uptheta}$ up to 160 mrad, though the plots only show up to 120 mrad for clearer visualization. For most simulations presented here, 9 iterations were used for the entire tilt range sampled (giving 1.86 mrad sample spacing) and an additional iteration was performed for the inner 120 mrad $\times$ 120 mrad (giving 0.94 mrad sample spacing). The temperature-dependent library used for Section~\ref{subsec:photoinducedMatching} was only computed up to 20 nm film thickness, so a 3.72 mrad sample spacing for the whole range and 1.86 mrad spacing for the inner quarter was sufficient.

This algorithm is robust to the complex variations of the diffraction intensities with tilt angle. Especially convenient is the hierarchical nature of this algorithm: Each sampling grid can also be used to calculate tilt-averaged diffraction patterns for smaller tilt spreads, ie. the grid used to compute the iteration $N$ for $\sigma_{\uptheta}$ is the same to compute the iteration $N-1$ for $\frac{\sigma_{\uptheta}}{2}$. This property allows us to compute a library of tilt-averaged patterns with varying $\sigma_{\uptheta}$ largely in parallel, with additional iterations applied to the successively smaller tilt spreads once the larger tilt spread calculations are converged. 

For kinematical and Bloch wave simulations, sample tilt is incorporated into the excitation error coefficients $s_{hkl}$. For the Bloch wave simulations, this also affects which Fourier components are included in the simulation at each tilt angle, as different beams are brought near their Bragg condition and contribute to the scattering process.

For the multislice calculations, we implement sample tilt by applying the Fourier shear theorem to the propagation operator, as follows:\cite{Kirkland_2010}

\begin{multline}
    \psi_{j+1}(\mathbf{q})= \\ \psi_{j}(\mathbf{q})\exp\left( -i\pi \Delta t \left[\lambda|\mathbf{q}|^{2} + 2 (\tan(\theta_{\rm{x}})q_{\rm{x}}+\tan(\theta_{\rm{y}})q_{\rm{y}})\right]\right)
\end{multline}

This allows to sample arbitrary tilt angles without changing the electrostatic potential slices. Prior works have suggested that this approximation can introduce significant error at angles beyond 1 degree,\cite{Kirkland_2010} while freestanding films studied in UED sometimes have more than 5 degrees RMS tilt spread. So, while this approximation is invoked here, future works could seek to implement approaches that improve accuracy for larger tilt spreads.\cite{ishizuka1982tiltedmultislice,ishizuka2004silveranniv}

\begin{figure*}
    \centering
    \includegraphics[width=17.5cm]{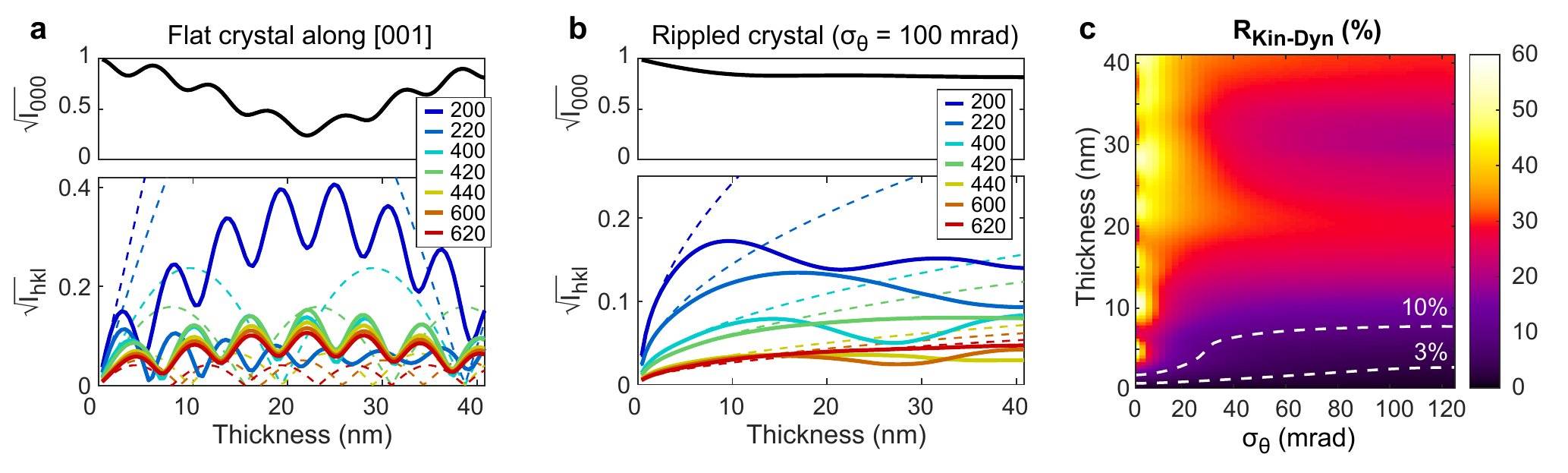}
    \caption{Role of dynamical scattering in relativistic (750 keV) ultrafast electron diffraction of oriented single crystal gold films. a) Primary ($I_{000}$) and diffracted ($I_{hkl}$) intensities for the [001] orientation as a function of thickness, computed using Bloch waves (solid lines) and kinematical theory (dashed lines). b) The same but averaged over a beam-sample orientation distribution with RMS tilt spread $\sigma_{\uptheta}$ = 100 mrad. c) R factor between kinematical and dynamical scattering calculations of the first seven diffraction orders ($R_{\rm{Kin-Dyn}}$) mapped to illustrate the difference between the models over varying film thickness and RMS tilt spread, $\sigma_{\uptheta}$. The white dashed contour lines mark film thicknesses beyond which R factor exceeds the noted value.}    \label{fig:figure1}
\end{figure*}

\subsection{Convergence}
\label{subsec:convergence}

To achieve good quantitative precision, some simulation parameters needed to be tuned until the diffraction signals converge. We used the crystallographic R factor as the metric, given by:
\begin{equation}
    R = \frac{\sum_{hkl} 
    \left|\sqrt{I_{hkl}^{i}} - \sqrt{I_{hkl}^{i_{\rm{max}}}}\right|}{\sum_{hkl} \sqrt{I_{hkl}^{i_{\rm{max}}}}}
\label{eq:Rfactor}
\end{equation}
which compares iteration $i$ and the final iteration, $i_{\rm{max}}$. All simulations in this work are converged until the R factor computed over the first seven diffracted orders (the ones we are interested in quantifying) is less than 1\%. For Bloch wave calculations, the parameters to converge are the thresholds that determine which diffracted beams are included, ie. the $q_{\rm{xy,max}}$ and $s_{\rm{max}}$ used here. For multislice calculations, the main parameter to converge is the real-space pixel size (q range), which must be sufficiently small (large) to include enough of the scattered beams for accurate diffraction calculations. The image dimensions are chosen to be powers of 2 for optimal speed of the fast Fourier transforms. Additionally, for all orientation-averaged simulations, the sampling density in the orientation space was increased until R < 1\% was achieved over the entire thickness and RMS tilt range studied.

\subsection{Programs}

We have created a MATLAB code library to perform all the calculations shown in this work, which is available online (see \hyperref[sec:codeavailability]{Code and Data Availability}). All simulation methods can be performed using CPUs, but for the multislice calculations in this work the time cost was prohibitive when performing the orientation-averaged calculations. As such, for the multislice calculations we have also implemented a version for GPU computing using ``gpuArray'' objects in MATLAB. 

Our main goal with these codes was to demonstrate and compare the accuracy of the described approaches for the UED quantification shown, so speed was not fully optimized. For reference, each orientation-averaged diffraction library (thickness from 0 to 40 nm and $\sigma_{\uptheta}$ from 0 to 160 mrad) took about 2.5 hours to compute using our multislice program on an NVidia Quadro K5000 GPU and about 1.7 hours using our Bloch wave program on an Intel iCore i7-8550U CPU. This time was reduced to about 23 minutes for the libraries computed up to just 20 nm thickness for Section~\ref{subsec:photoinducedMatching} using Bloch waves. For all methods, the initial setup computations required several seconds, and then each orientation took less than a second to compute. Several approaches to improve performance have been demonstrated by others and could be implemented in the future. Examples for Bloch waves have included using off-diagonal matrix elements to compute an array of tilts simultaneously\cite{sinkler2010DynamicalPEDSims} and GPU acceleration.\cite{koch2014GPUAcceleratedBW} We also note that open-source, high-performance programs for Bloch wave and multislice simulations have been developed by others,\cite{lobato2015multem,Ophus_2017} though some adaptation may be needed to suit UED simulation.

\section{Role of dynamical scattering in UED of single-crystal foils}
\label{sec:Sims}

\subsection{Static diffraction peak signals in flat and rippled foils}

The importance of dynamical scattering in oriented single-crystal foils is evident in the complex evolution of the electron wave through nanometer-scale ultrathin films even at relativistic beam energies. Here, we show diffraction simulations computed for a 750 keV electron wave passing through gold films up to 40 nm thick oriented along [001] with mean square displacements of $\overline{u^2} = 0.024 $ \AA$^2$ (This $\overline{u^2}$ has been measured by others for films at 300 K using X-ray diffraction).\cite{owenwilliams1947DWFGold,singh1971DWFmetals}  Figure~\ref{fig:overview}b shows channeling plots calculated using the multislice method, which illustrate the amplitude and phase of an electron plane wave passing through a flat film. Whereas kinematical scattering approximations assume the material is a ``weak phase object,'' the strong and dense gold atomic columns locally shift the electron phase by more than $\pi$ within just a few unit cells. This imparts dramatic modifications of the electron wave amplitude, including significant channeling along the atomic columns within a few nm. This in turn leads to complex, oscillatory thickness dependence of the primary and diffracted beam intensities as shown in Figure~\ref{fig:figure1}a. Within just a few nm, the intensities deviate from those predicted by kinematical theory (dashed lines) despite the relativistic electron beam energy. 

When averaging over a large beam-sample orientation distribution, much of the oscillatory behavior is smoothed out; However, the diffracted intensities still show a complex behavior as a function of thickness that is not captured by kinematical theory. Figure~\ref{fig:figure1}b shows the total diffracted intensities calculated for a Gaussian tilt distribution with a large $\sigma_{\uptheta}$ = 100 mrad as might be found in a strongly rippled thin-film sample, which nonetheless shows significant variation in the diffracted intensities within this thickness range. 

The deviation of dynamical scattering models from kinematical theory can be quantified by computing the crystallographic R factor between the diffracted intensities obtained using two methods ($I_{hkl,1}$ and $I_{hkl,2}$):
\begin{equation}
    R_{1-2} = \frac{\sum_{hkl} 
    \left|\alpha\sqrt{I_{hkl,1}} - \sqrt{I_{hkl,2}}\right|}{\sum_{hkl} \sqrt{I_{hkl,2}}}
\label{eq:Rfactor2}
\end{equation}
Where the scaling factor $\alpha$ is fit to minimize R. The R factors computed between kinematical and dynamical simulations, $R_{\rm{Kin-Dyn}}$, computed using the first seven diffracted orders for films of varying thickness and tilt spread are shown in Figure~\ref{fig:figure1}c. Larger tilt spread increases sampling away from the zone axis, where channeling and multiple scattering effects are reduced, and averaging over a broad range smooths out these effects. Thus, larger tilt spread increases the range of film thickness for which the diffracted intensities can be approximately calculated using kinematical theory. Still, in this case, $R_{\rm{Kin-Dyn}}$ > 10 \% is observed for films thicker than 8 nm even at large, $\approx$ 100 mrad RMS tilt spreads. In many UED experiments, RMS tilt spreads can be much smaller, especially if films are prepared on sturdy membrane supports, and deviations from kinematical theory are more pronounced.\cite{TaTe2UED}

Notably, the large variations observed occur well within the elastic mean free path, calculated using the classic formula like in Ref.~\citenum{carterwilliamsTEMbook} to be 18.8 nm for 750 keV electrons through gold. However, the mean free path essentially considers an average electrostatic potential where scattering events are uncorrelated, while an oriented single crystal presents highly correlated scattering events along the atomic columns which more rapidly lead to strong multiple scattering and violations of kinematical approximations. This observation highlights the need to apply dynamical scattering models to UED signals from single-crystal foils at thicknesses well below the elastic mean free path.

\subsection{Dependence on electron probe energy}

\begin{figure}[h]
    \centering
    \includegraphics[width=8.5cm]{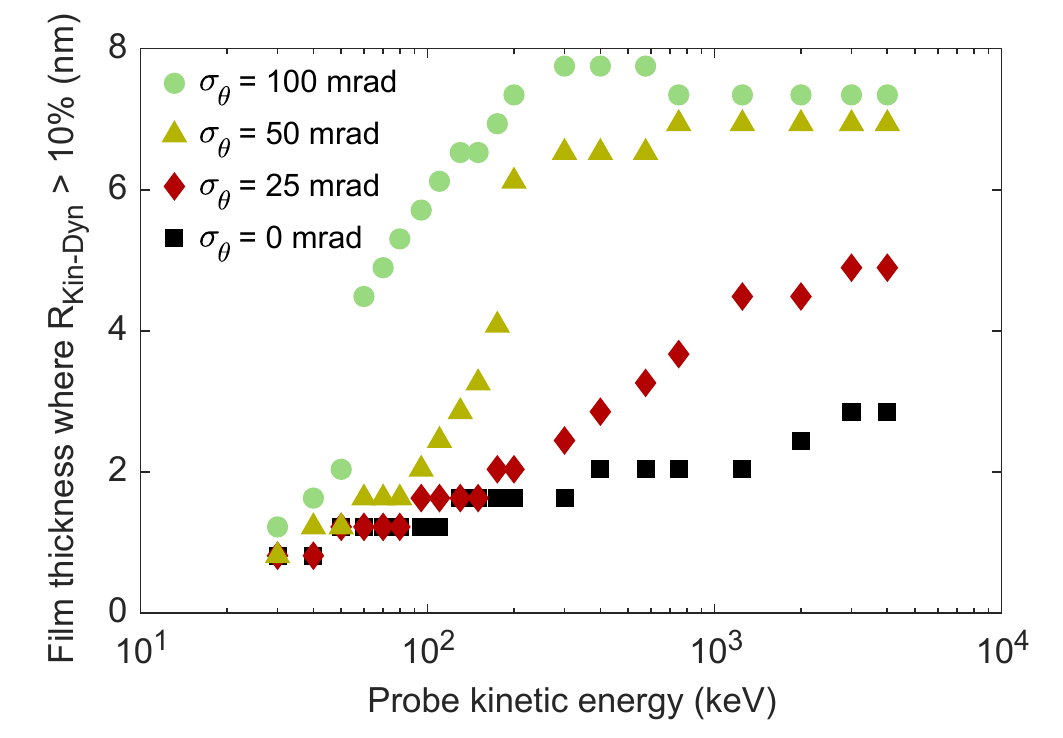}
    \caption{Dependence of the onset of dynamical scattering effects in single-crystal gold films on electron probe energy. The film thickness where $R_{\rm{Kin-Dyn}}$, the R factor computed between kinematical and multislice calculations of the first seven diffraction orders, first exceeds 10\% is plotted as a metric for four RMS tilt spreads ($\sigma_{\uptheta}$).}
    \label{fig:energyDep}
\end{figure}

One of the motivations for developing and utilizing UED beamlines with higher electron probe energy is to reduce dynamical scattering effects like those shown in the previous section. Here, we examine how electron probe energy affects the validity of kinematical approximations for the case of single-crystal gold films. To do this, we performed multislice simulations for the same film thicknesses and beam-sample orientation distributions chosen previously, but now for various probe energies between 30 keV and 4 MeV. We then extracted the film thickness at which $R_{\rm{Kin-Dyn}}$ first exceeds 10\% for each $\sigma_{\uptheta}$ and beam energy. We note that the 10\% value was chosen as an example and the true acceptable range for kinematical approximation depends on many factors including the structural parameters being quantified and the desired accuracy. Still, the results for selected $\sigma_{\uptheta}$ shown in Figure~\ref{fig:energyDep} highlight some interesting trends and provide a sense of scale. 

Notably, we observe different behaviors depending on the beam-sample orientation distribution. For flat foils and low-divergence probes ($\sigma_{\uptheta}$ near zero), the acceptable thickness range gradually increases with the beam energy, though still limited to less than 3 nm even at 4 MeV. For highly rippled foils where $\sigma_{\uptheta}$ is tens to hundreds of mrad, we observe an initial increase in acceptable thickness range and then a plateau at about 7.5 nm beyond a threshold energy. The threshold energy appears to decrease for broader orientation distributions. This behavior can be rationalized as follows. In the flat case, strong oscillations dominate (see Figure~\ref{fig:figure1}a) with periods set by the extinction distances which increase with beam energy even at relativistic energies (analogous to the ones defined in Equation~\ref{eq:extDist}. Meanwhile, orientation averaging smooths out this oscillatory behavior (see Figure~\ref{fig:figure1}b) and the deviation from kinematical validity is instead set by the interaction parameter $\sigma_{\rm{e}}$, which levels off for relativistic beam energies.\cite{Kirkland_2010} Altogether, these results show that using higher probe energies and orientation averaging can help to extend the validity of kinematical approximations, but only up to a point; in many cases, dynamical scattering models will still be required to accurately match the signals.

\begin{figure*}
    \centering
    \includegraphics[width=17.5cm]{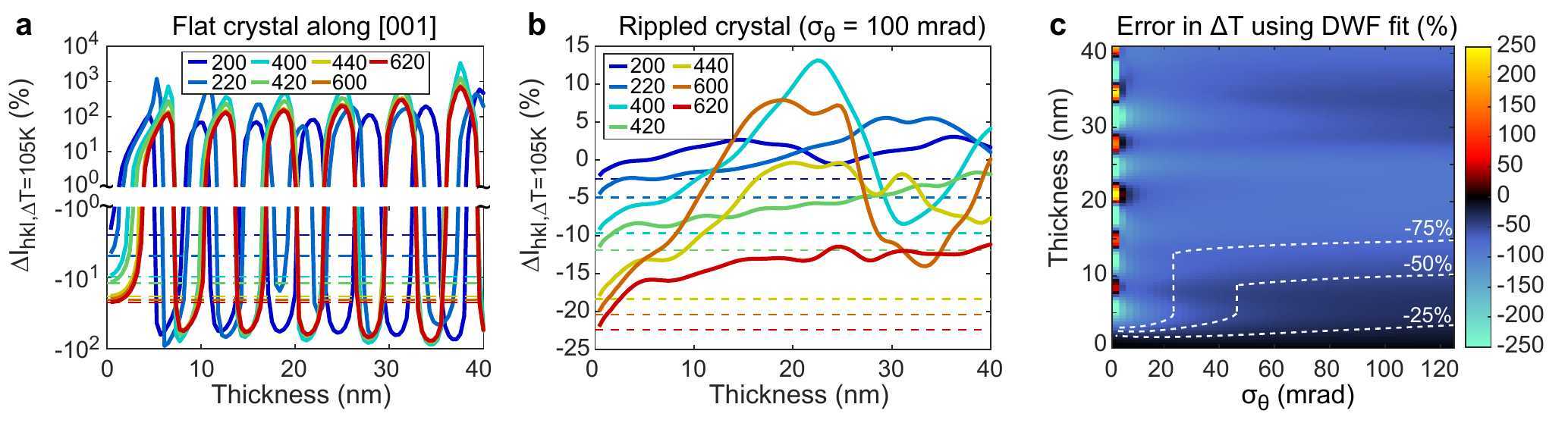}
    \caption{Role of dynamical scattering in lattice temperature effects and quantification. a) Simulated change in diffracted intensities ($\Delta I_{hkl}$) from a flat, [001]-oriented gold single-crystal film for a temperature rise $\Delta T$ = 105 K ($\Delta \overline{u^2} = 0.008$ \AA$^2$), calculated using Bloch waves (solid) and kinematical theory (dashed lines). b) The same, but for a beam-sample orientation distribution with RMS tilt spread $\sigma_{\uptheta}$ = 100 mrad. c) Percent error in extracting $\Delta T$ from the Bloch wave model diffracted intensity changes by fitting the Debye-Waller factor (DWF). The white dashed contour lines mark thickness thresholds beyond which the error first exceeds the noted value.}
    \label{fig:2}
\end{figure*}

\subsection{Temperature dependence of diffraction signals}

Dynamical scattering not only affects the individual peak intensities, but also how they change with structural parameters such as the lattice temperature. Figure~\ref{fig:2} compares the diffraction peak intensity changes computed for a 105 K temperature increase ($\Delta \overline{u^2}$ = 0.008 \AA$^2$) using Bloch wave and kinematical scattering calculations. Whereas kinematical theory predicts the peak intensities will be scaled by the Debye-Waller factor regardless of crystal thickness, Bloch wave calculations show significant deviations from this prediction within few-nm film thicknesses. This is especially apparent for a flat crystal oriented on zone (Figure~\ref{fig:2}a) where diffracted intensities can be nearly eliminated or dramatically enhanced (up to and exceeding a factor of 10) with the temperature change depending on the film thickness. Orientation averaging again smooths the variations, but there are still significant deviations within several nm thickness. 

Importantly, as the film thickness in the rippled case increases, the mean absolute intensity change due to temperature tends to zero, with some diffraction peaks gaining intensity and others losing intensity as temperature increases in films thicker than $\approx$ 10 nm. This behavior is in stark contrast to kinematical theory, where all diffraction peaks lose intensity with increasing temperature. This can be understood by considering that in the regime of strong multiple scattering, electrons are scattering back and forth between various diffracted beams and the primary beam, so an increase in temperature merely modifies the distribution of these multiply-scattered electrons throughout the various beams. Indeed in the flat film, this even leads to conditions where diffracted beams are observed to increase in intensity on average with increasing temperature. 

As a result, fitting the Debye-Waller factor (DWF) to quantify lattice temperature changes in the dynamical scattering regime can give large errors, even if fitting several diffraction orders. The error in $\Delta T$ extracted by a least-squares DWF fit (Equation~\ref{eq:DWFitting}) to the intensity changes for the first seven diffraction orders calculated using the Bloch wave model for $\Delta T$ = 105 K is shown in Figure~\ref{fig:2}c. Again, orientation averaging reduces oscillatory behaviors and smooths out the error, extending the range of validity for DWF fitting compared to flat films, but significant errors still emerge within several nm. Notably, DWF fitting tends to significantly underestimate $\Delta T$ in these rippled film models due to the trend towards zero mean intensity change observed in panel b, even extracting nearly zero temperature change in rippled films near 20 nm thickness (error $\approx$ -100\%). The deviations worsen for smaller tilt spreads, with DWF fitting massively over- or underestimating the temperature rise, at some thicknesses even extracting a temperature decrease instead of an increase (error < -100 \%). The fitting results can also vary dramatically depending on which diffraction orders are included in the analysis and how they are weighted.

Altogether, these simulations illustrate the important role of dynamical scattering in UED of flat and rippled single-crystal foils even at relativistic beam energies, using gold as an example. They also reinforce that kinematical scattering models are insufficient for quantitative matching and analysis in films with strong multiple scattering. In the next section, we demonstrate that the dynamical scattering models shown here can be used in practice to quantitatively match experimental UED data and retrieve structural parameters.

\section{Quantification of photoinduced lattice temperature rise in single-crystal gold films}

Here, we apply the described scattering models to quantitatively match and analyze a UED experiment performed on a single-crystal gold foil at the HiRES beamline. A 750 keV electron probe was used with a 150 $\rm{\mu}$m RMS spot size. The [001]-oriented freestanding single-crystal gold foil on a 3 mm diameter, 300 mesh TEM grid was purchased from Ted Pella, and was quoted to be 11 nm thick.

\begin{figure*}
    \centering
    \includegraphics[width=17.5cm]{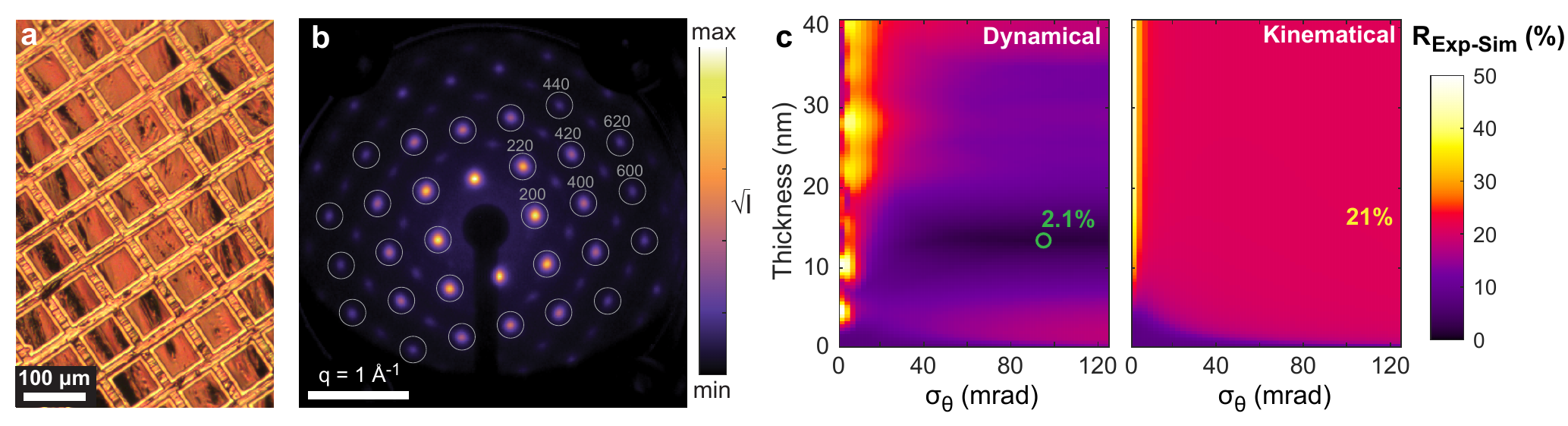}
    \caption{Quantitative matching of diffraction simulations to a UED pattern recorded at the HiRES beamline. \textbf{a} Optical micrograph of the [001]-oriented single-crystal foil, showing the rippled topography. \textbf{b} Experimental UED pattern recorded using 750 keV electrons, labeled with the diffraction orders studied. \textbf{c} R factor between measured and simulated signals ($R_{\rm{Exp-Sim}}$) using Bloch waves (dynamical) and kinematical models over a range of crystal thickness and RMS tilt spread ($\sigma_{\uptheta}$) with $\overline{u^2}$ = 0.024 \AA$^2$. The green circle marks the best fit using the dynamical model, where $R_{\rm{Exp-Sim}}=2.1\%$. At the same thickness and tilt spread, kinematical theory gives $R_{\rm{Exp-Sim}}=21\%$.}
    \label{fig:MS_staticMatchingGold}
\end{figure*}

\subsection{Matching an experimental UED pattern}

We first applied the dynamical scattering models to match a UED pattern recorded from the sample without any optical excitation. An optical micrograph of the film is shown in Figure~\ref{fig:MS_staticMatchingGold}a. Large rippling is evident in the freestanding gold foil. The experimental UED pattern recorded at HiRES is shown in Figure~\ref{fig:MS_staticMatchingGold}b. The peak positions in reciprocal space are consistent with those expected for the [001] orientation of gold. The apparent four-fold symmetry of the pattern indicates the film is well oriented along the zone axis on average, and suggests that the orientation distribution can be reasonably approximated as isotropic. We also note the peak widths are dominated by the electron probe divergence $\sigma_{\theta,\rm{probe}}$, measured to be about 0.33 mrad. In this case, the total $\sigma_{\theta}$ of the beam-sample orientation distribution will be dominated by the large sample rippling.

All experimentally recorded diffraction patterns examined in this work were obtained by the following method. 30 sub-frames were recorded for each pattern, and a dark current reference (recorded with both the photocathode and pump lasers off) was subtracted from them. Then, the alpha-trimmed mean of the sub-frames was computed, removing outlier x-ray spikes. The peak intensities for the first seven diffraction orders were extracted by fitting radially symmetric 2D Gaussians of the form:
\begin{equation}
    I(\mathbf{q}-\mathbf{q_{hkl}}) = I_{hkl}\exp\left(-\frac{(\mathbf{q}-\mathbf{q_{hkl}})^2}{2\sigma_q^2}\right)+k_{hkl}
\end{equation}
Where $\mathbf{q_{hkl}}$ is the peak location in reciprocal space, $I_{hkl}$ is the peak intensity, $k_{hkl}$ is the (uniform) background level beneath the peak, and $\sigma_q$ is the peak width. $I_{hkl}$ is the parameter of interest to fit for each peak in each pattern. $q_{hkl}$ is also refined for all peaks in each pattern to accommodate drift and thermal expansion, and $k_{hkl}$ is likewise refined to accommodate background fluctuations and underlying thermal diffuse background. Meanwhile, $\sigma_q$ is fixed to the 200 peaks in the first laser off pattern, then fixed to this value for all peaks in all subsequent patterns; This is done because the peak width in our experiment is dominated by the angular spread of the probe. Only those peaks with a corresponding Friedel pair visible in the pattern were included to reduce error from slight misorientation relative to the zone axis.

The agreement between the experimental intensities and the diffraction patterns simulated with $\overline{u^2} = 0.024$ \AA$^2$ for varying film thickness and tilt spread was quantified by computing $R_{\rm{Exp-Sim}}$ (Eq.~\ref{eq:Rfactor2}) using both dynamical and kinematical models, displayed in Figure~\ref{fig:MS_staticMatchingGold}c. Remarkably, the dynamical scattering models achieve a tenfold reduction in $R_{\rm{Exp-Sim}}$, yielding 2.1\% at the optimal thickness and RMS tilt spread compared to 21\% obtained with kinematical models. Furthermore, the optimum parameters are physically reasonable: a thickness of 13.5 nm is in good agreement with the 11 nm quoted by the vendor, and the large 95 mrad RMS tilt spread is reasonable given the optically visible wrinkling. Both models are only weakly dependent on tilt spread beyond $\approx$ 20 mrad RMS, so in this range the precise value of tilt spread is less important; on the other hand, the dynamical scattering model provides a precise determination of the film thickness that the kinematical model cannot. 

\begin{figure*}
    \centering
    \includegraphics[width=17.5cm]{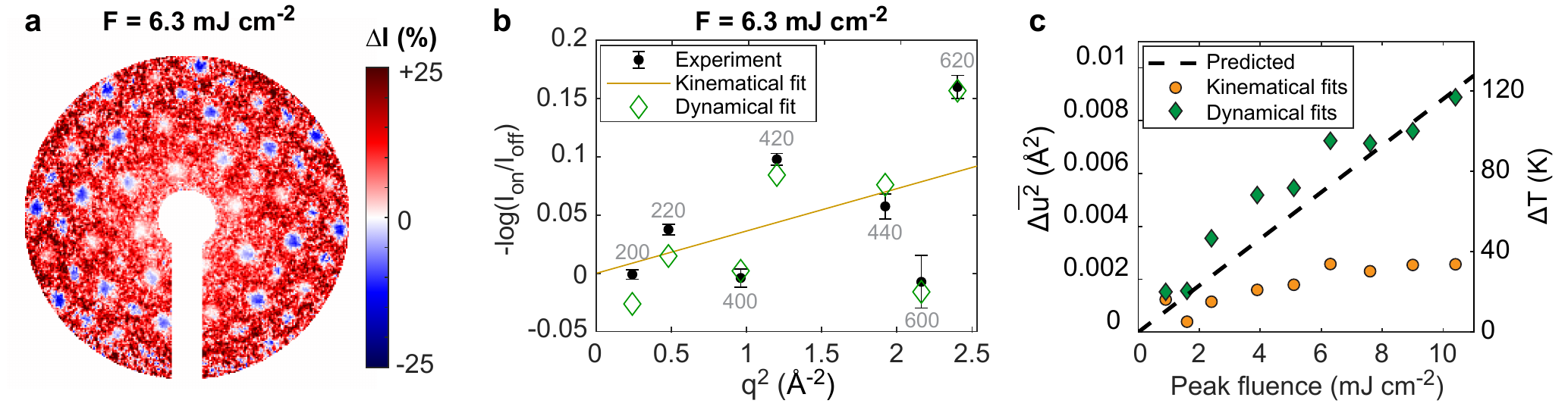}
    \caption{Quantifying light-induced lattice heating from pump-probe UED measurements. \textbf{a} Example of a photoinduced difference pattern recorded at HiRES using a peak laser fluence of 6.3 mJ cm\textsuperscript{-2}. \textbf{b} Measured ratio of laser-on to laser-off diffraction intensities $I_{\rm{on}}/I_{\rm{off}}$ (black dots), Debye-Waller factor fit (orange line), and Bloch wave model fit (green diamonds). \textbf{c} Extracted fluence-dependent changes in total RMS atomic displacements ($\Delta \overline{u^2}$) and lattice temperature ($\Delta T$) using Debye-Waller factor fitting (Kinematical) and Bloch wave calculations (Dynamical). Changes predicted using known optical constants of gold are superimposed as a black dashed line for comparison.}
    \label{fig:MS_photoinducedTempRise}
\end{figure*}

\subsection{Matching photoinduced difference patterns}
\label{subsec:photoinducedMatching}

Next, we applied the simulations to retrieve the photoinduced lattice temperature from a pump-probe UED measurement.  Photoexcited UED patterns from the same single-crystal gold film were measured using $\lambda$ = 1030 nm pump laser pulses at a 0.5 kHz repetition rate with varying fluence. At each fluence, UED patterns were recorded as the pump-probe delay, $\Delta t$, was scanned from -17.3 to +56.0 ps using 6.67 ps steps. A coarse sampling was used in these measurements with a focus on extracting fluence-dependent temperature rise rather than the fine temporal dynamics. The average difference pattern recorded after the arrival of laser pulses with 6.3 mJ cm\textsuperscript{-2} (from +22.5 to +56 ps) is shown in Figure~\ref{fig:MS_photoinducedTempRise}a as an example. The coherent Bragg diffraction peaks are generally suppressed and the diffuse scattering background generally increases as expected for an increase in incoherent thermal motions. However, the diffraction peak intensity changes deviate from the scaling of the Debye-Waller factor: For instance, the 200, 400, and 600 peaks show little change while the 220, 420, and 620 peaks show large changes. 

The time-dependent diffraction peak intensities were extracted from the UED datasets with Gaussian peak fitting, and average peak intensities before and after $\Delta t = 0$ were calculated. The change in mean square atomic displacements was then retrieved before and after time zero both by fitting the Debye-Waller factor (Equation~\ref{eq:DWFitting}) and the Bloch wave models to the intensity changes measured for the first seven diffracted orders. In the Bloch wave approach, the thickness and RMS tilt spread of the film were fixed, and $\Delta \overline{u^2}$ was optimized by interpolating the peak intensities between calculations performed at 15 values of $\overline{u^2}$ ranging from $0.024$ \AA$^2$ to $0.038$ \AA$^2$. Both kinematical and Bloch wave models are fit by minimizing the mean square error of $-\log(\frac{I_{hkl}}{I_{0,hkl}})$ for the first seven diffracted orders. 

The results of this procedure for the after time zero diffraction intensities for a peak fluence of 6.3 mJ cm$^{-2}$ are illustrated by Figure~\ref{fig:MS_photoinducedTempRise}b. Indeed, the intensity changes predicted using Bloch waves (green diamonds) are a better match to the observations (black circles) than are those predicted by the Debye-Waller factor (dashed line), reducing the least-squares residual by about a factor of 3. The variations in intensity change between orders are largely captured by the dynamical scattering models used here, though differences still remain, perhaps due to differences between the simulated and actual orientation distribution of the sample or to finer details not yet accounted for such as inelastic scattering effects. Photoinduced strain could also alter the orientation distribution of the sample and contribute to peak intensity changes; that said, we note that including $\Delta\sigma_{\theta}$ as a second fit parameter (done in a separate analysis) does not have a strong systematic effect on the results, with $\Delta\sigma_{\theta}$ less than 3 mrad and $\Delta u^2$ modified by about $\pm$10\% on average for all the peak fluences studied. 

The photoinduced change in mean square displacements and lattice temperature rise are plotted as a function of peak fluence in Figure~\ref{fig:MS_photoinducedTempRise}c. The relationship between mean square displacements and lattice temperature rise in gold is approximately linear in the studied range of lattice temperatures (about 3.947 $\times$ 10\textsuperscript{4} K/\AA\textsuperscript{2}).\cite{owenwilliams1947DWFGold,singh1971DWFmetals} Strikingly, the dynamical scattering models retrieve photoinduced lattice temperatures that are more than three times higher than those retrieved using the Debye-Waller factor approach. 

Comparing to estimations of lattice temperature rise for the given peak fluence using the known optical constants of gold supports the accuracy of the dynamical scattering models. The lattice temperature rise $\Delta T = T_{\rm{f}}-T_{\rm{i}}$ was calculated by relating the absorbed energy density (ie. in J/mol) $U_{\rm{abs}}$ to the heat capacity of the material $C_{\rm{p}}$:
\begin{equation}
    U_{\rm{abs}} = F_{\rm{inc}}A\frac{V_{\rm{mol}}}{t} =  \int_{T_{\rm{i}}}^{T_{\rm{f}}} C_{\rm{p}}(T)
\end{equation}
where $F_{\rm{inc}}$ is the incident laser fluence, $A$ is the absorbance of the material at the incident photon energy, $V_{\rm{mol}}$ is the molar volume, and $t$ is the thickness of the film. The absorbance in the 13.5 nm film of $\lambda$ = 1030 nm light, using n = 0.153 and k = 6.654,\cite{olmon2012dieleAul} was calculated using the coherent transfer matrix method\cite{byrnes2016tmm} to be 3.7\%. Using this and the measured temperature-dependent heat capacity of gold,\cite{takahashi1986CpGold} the temperature rise per unit of incident laser fluence was found to be 11.6 K / (mJ cm\textsuperscript{-2}). Fitting a line to the temperatures retrieved with Bloch waves gives a slope close to this of 12.3 K / (mJ cm\textsuperscript{-2}) whereas the kinematical approach gives 4.0 K / (mJ cm\textsuperscript{-2}).

\section{Conclusions}

This work demonstrates the importance and application of dynamical scattering theory for quantitative analysis of ultrafast electron diffraction patterns. As shown here for single-crystal gold foils, diffraction signals are influenced by film thickness, temperature, and topography in ways that are sometimes entirely opposite of intuitions from kinematical theory. By virtue of the proposed modified treatment we are able to reach accurate UED pattern matching and lattice temperature quantification in a single-crystal experiment. We also show how a kinematical approach to the same problem would lead to greatly underestimated lattice temperatures. 

The described models can be further extended to a wide range of experiments and samples. For instance, they are readily extended to multilayered single-crystal films by simulating each layer in series, and can be extended for crystals of any space group. Larger simulation cells and complex symmetries may demand higher performance programs and computing resources for practical computation and refinements. Other physical parameters not included here may also be important for accurate quantification for different classes of specimen and different experimental setups. Some possible examples include anisotropic thermal displacement parameters, thermal diffuse scattering, core losses, and coherence properties of the probe.

Other structural parameters besides the lattice temperature can be refined. For instance, crystallographic parameters like Wyckoff positions can be used as variables, and refining these in combination with the thermal displacements could be used to quantify and separate simultaneous crystal structure change and thermal motions during structural phase transformations. 

More widespread availability and use of dynamical scattering models for UED will enable more detailed information to be retrieved from UED experiments and will expand the technique's capabilities and scientific breadth. A few examples of materials that could become more accessible include thick, multilayered crystals such as large epitaxial superlattices; single-crystal nanowires or nanoparticles with critical dimensions in the dynamical scattering regime; and buried layers in thick semiconductor device stacks. In the long term, improving quantitative matching of UED patterns could ultimately enable full crystal structure refinement of transient structures, such as photoinduced nonequilibrium phases. To recover complete 3D movies of the atomic coordinates and thermal motions in single crystals, for instance from UED tilt series, would mark a major milestone for UED and provide detailed structural knowledge of transient intermediates, metastable phases, coherent lattice responses, and the overall energy flow and structural dynamics.

\begin{appendix}
\label{appendix}

\section*{Appendix: Comparison of Bloch wave and multislice methods}

\begin{figure}[h]
    \centering
    \includegraphics[width=8.5cm]{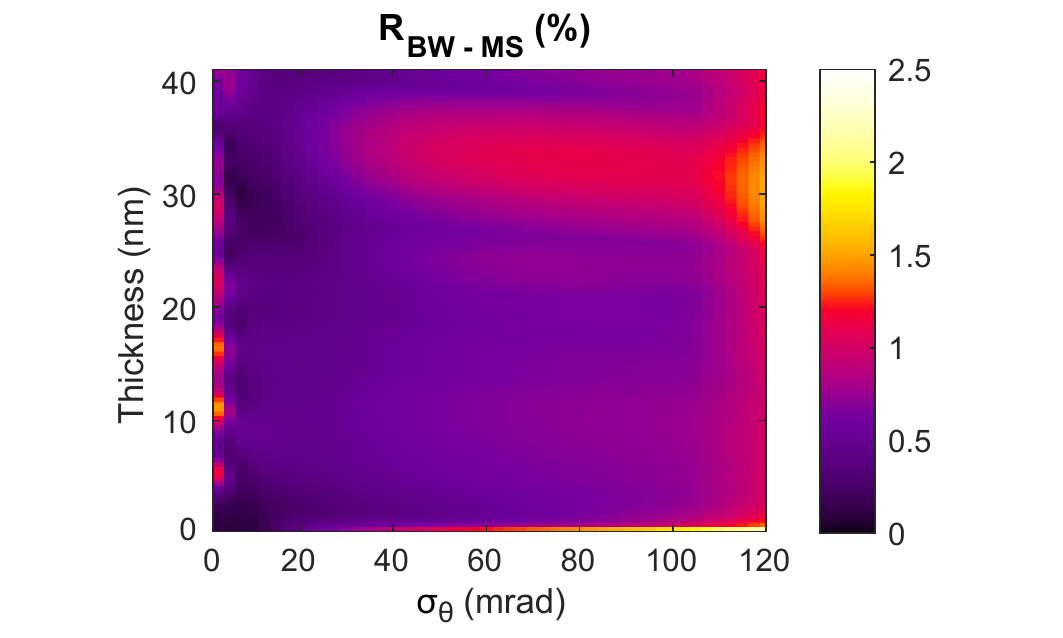}
    \caption{R factor between Bloch wave and multislice calculations ($R_{\rm{BW-MS}}$) of the first seven diffraction orders for the conditions in Figure~\ref{fig:figure1}.}
    \label{fig:BWMS}
\end{figure}

In principle, Bloch wave and multislice approaches can be equivalent since they are both methods of solving the same Schr\"{o}dinger equation for fast electrons through the specimen under nearly the same approximations. In practice, however, which method is more convenient, faster to compute, or more accurate depends on the specimen material and geometry, experimental conditions, and the signals being modeled. In our case of modeling flat and rippled single-crystal gold Bragg diffraction peaks, we find that the approaches give nearly identical results over most of the thickness and tilt spread range studied. The R factor between the diffraction signals calculated using Bloch wave and multislice models, $R_{\rm{BW-MS}}$, is shown in Figure~\ref{fig:BWMS}. We find $R_{\rm{BW-MS}}$ < 1\% for most of the range studied, though errors slightly increase for $\sigma_{\uptheta}$ > 100 mrad perhaps due to limitations of the approximate tilt correction to the multislice calculation discussed in Section~\ref{subsec:avgrippling}.

For the small unit cell and low q resolution of these simulations, we find our Bloch wave model is several times faster than our multislice model. For more complex, larger simulation cells with higher q resolution, multislice may become the faster approach due to more favorable scaling with the number of reciprocal space points.\cite{Kirkland_2010}

\end{appendix}

\begin{acknowledgments}

D.B.D. and A.M.M. acknowledge support from STROBE: A National Science Foundation Science and Technology Center under Grant No. DMR 1548924. C.O. acknowledges support from the DOE Early Career Research Award program. K.M.S. and D.F. acknowledge support from the Laboratory Directed Research and Development (LDRD) Program of Lawrence Berkeley National Lab under U.S. Department of Energy (DOE) Contract No. DE-AC02-05CH11231. Development and operation of the HiRES instrument (D.F.) was supported by DOE Contract No. DE-AC02-05CH11231. Work at the Molecular Foundry was supported by the Office of Science, Office of Basic Energy Sciences, of the U.S. Department of Energy under Contract No. DE-AC02-05CH11231. 

\end{acknowledgments}

\section*{Code and data availability}
\label{sec:codeavailability}
The programs and data that support the findings of this study are openly available in GitHub at \url{https://github.com/dbdurham/QuantUEDSim}.\cite{quantUEDSim}

\appendix

\bibliography{main}

\end{document}